\documentclass[prb,twocolumn,showpacs]{revtex4}
\usepackage{graphicx}
\usepackage{amsmath}
\usepackage{amssymb}
\usepackage{color}

\newcommand{\be}{\begin{equation}}
\newcommand{\ee}{\end{equation}}

\begin{document}

\title{Anomalous growth of thermoelectric power in gapped graphene}
\author{S.G.~Sharapov}
%\email{sharapov@bitp.kiev.ua}
\affiliation{Bogolyubov Institute for Theoretical Physics, National Academy of Science of
Ukraine, 14-b Metrologicheskaya Street, Kiev, 03680, Ukraine}
\affiliation{Mediterranean Institute for Fundamental Physics, Rome, Italy}
\author{A.A.~Varlamov}
\affiliation{CNR-SPIN, University ``Tor Vergata'', Viale del Politecnico 1, I-00133 Rome, Italy}
\affiliation{Mediterranean Institute for Fundamental Physics, Rome, Italy}
\date{\today }

\begin{abstract}
There exist experiments indicating that at certain conditions, such as an
appropriate substrate, a gap of the order of 10 meV can be opened at the
Dirac points of a quasiparticle spectrum of graphene. We demonstrate that
the opening of such a gap can result in the appearance of a fingerprint bump of the
Seebeck signal when the chemical potential approaches the gap edge. The
magnitude of the bump can be up to one order higher than the already large value
of the thermopower occurring in graphene. Such a giant effect, accompanied
by the nonmonotonous dependence on the chemical potential, is related to
the emergence of a new channel of quasiparticle scattering from impurities
with the relaxation time strongly dependent on the energy. We analyze
the behavior of conductivity and thermopower in such a system, accounting for
quasiparticle scattering from impurities with the model potential in a
self-consistent scheme. Reproducing the existing results for the case of
gapless graphene, we demonstrate a failure of the simple Mott formula in the
case under consideration.
\end{abstract}

\maketitle

%\email{}

%81.05.ue 	Graphene (for structure of graphene, see 61.48.Gh; for phonons in graphene, see 63.22.Rc; for thermal properties, see %65.80.Ck; for graphene films, see 68.65.Pq; for electronic transport, see 72.80.Vp; for electronic structure, see 73.22.Pr; for %optical properties, see 78.67.Wj)

%65.80.Ck 	Thermal properties of graphene

% 72.80.Vp 	Electronic transport in graphene

%\keywords{GRAPHENE}

Control of heat flows and minimization of heat losses is an important
aspect of designing modern nanoelectronic devices, in particular those based
on graphene \cite{Geim2009Science}. Experiments indicate \cite%
{Grosse2011NatureNano} that the thermoelectric effect in graphene accounts for
up to one-third of the contact temperature changes and thus it can play a
significant role in cooling down of such systems. The measured thermopower $%
S $ reaches the value $k_{B}/e\sim 100\mu V/K$ at room temperatures, where $%
k_{B}$ is the Boltzmann constant and $-e<0$ is the electron charge. This and
other experimental results \cite%
{Wei2009PRL,Zuev2009PRL,Checkelsky2009PRB,Wang2011PRB} on thermoelectric
transport in graphene were understood theoretically, mostly basing them on a
simple Mott relation between thermopower and the logarithmic derivative of the
electrical conductivity $\sigma (\mu ,T)$,
\begin{equation}
S(\mu ,T)=-\frac{\pi ^{2}}{3e}T\frac{d}{d\mu }[\ln \sigma (\mu ,T=0)],
\label{Mott-formula}
\end{equation}%
where $\mu $ is the chemical potential and $T$ is the temperature
(we set $k_{B}=1$) \cite{xx}. However, the cited experiments show that the Mott
formula (\ref{Mott-formula}) fails when $\mu $ approaches the vicinity of
the Dirac point at high temperatures, especially in high-mobility graphene
\cite{Wang2011PRB}. The available theoretical analysis \cite%
{Lofwander2007PRB,Hwang2009PRB,Ugarte2011PRB} shows that the failure of
Eq.~(\ref{Mott-formula}) can be attributed to breaking of the conditions of its
applicability, which read as $T\ll |\mu |$ and/or $T\ll \gamma $
[here $\gamma $ is the characteristic energy scale on which the conductivity $%
\sigma (\mu ,T=0)$ varies around the Fermi level].

The purpose of the present paper is to show that the already large value of the
thermopower occurring in graphene can be further increased up to one order
of magnitude by opening in different ways (see, e.g., Refs.~\onlinecite%
{Gusynin2007IJMPB,Vozmediano2011}) a gap $\Delta $ in its quasiparticle
spectrum. We will show that such an opening is accompanied by the emergence of
a new channel of quasiparticle scattering with the relaxation time
strongly dependent on energy, and it results in the appearance of a giant bump
in thermopower when chemical potential approaches the gap edge. The
situation here turns out to be very similar to the well known anomaly of
thermopower close to the electronic topological transition
(see Refs.~\onlinecite{Varlamov1989AP,Blanter1994PR} for a review) related to 
the scattering of electrons from all of the extended periphery of the Fermi surface 
to the ``trap,'' presented by the small new void or narrow ``neck'' of the latter.

It is worth mentioning that some experimental findings \cite%
{Zhou2007NatMat,Li2009PRL} indicate that, indeed, a gap in the graphene
quasiparticle spectrum opens at the Dirac point, and probably it can be
attributed to the effect of the substrate. Yet, this issue still remains an
open problem of the physics of graphene \cite{Gusynin2007IJMPB,Vozmediano2011}.
This is why, in principle, one can apply the results obtained below 
and use thermopower measurements as a sensitive probe of the
gap opening in the graphene spectrum.

The Hamiltonian for graphene can be written down in the momentum
representation as
\begin{equation}
\widehat{H}=\sum_{\sigma }\int_{BZ}\frac{d^{2}p}{(2\pi )^{2}}\Upsilon
_{\sigma }^{\dagger }(\mathbf{p})[\widehat{\mathcal{H}}(\mathbf{p})-\mu
\widehat{\tau}_{0}]\Upsilon _{\sigma }(\mathbf{p}),  \label{Hamgraph}
\end{equation}%
where%
\begin{equation*}
\widehat{\mathcal{H}} (\mathbf{p})= \widehat{\tau}_{+} \phi \left( \mathbf{p}%
\right) + \widehat{\tau} _{-}\phi ^{\ast}\left( \mathbf{p}\right) +\Delta
\widehat{\tau}_{3},
\end{equation*}
$\widehat{\tau}_{0}$, $\widehat{\tau}_{3}$, and $\widehat{\tau}_{\pm }= (\widehat{%
\tau}_{1} \pm i \widehat{\tau}_{2})/2$ are Pauli matrices acting in the
sublattice space on the spinors, $\Upsilon _{\sigma }(\mathbf{p})$ and $%
\Upsilon _{\sigma }^{\dagger }(\mathbf{p})=\left( a_{\sigma }^{\dagger }(%
\mathbf{p}),b_{\sigma }^{\dagger }(\mathbf{p})\right) $, with the creation (annihilation)
operators of electrons $a_{\sigma }^{\dagger }(\mathbf{p})$, $b_{\sigma
}^{\dagger }(\mathbf{p})$ ($a_{\sigma }(\mathbf{p})$, $b_{\sigma}(\mathbf{p})$)
corresponding to $A$ and $B$ sublattices, respectively,
and spin subscript $\sigma$. The full form of the complex function $%
\phi (\mathbf{p})$ is provided, for example, in Ref.~\onlinecite{Gusynin2007IJMPB}
and in present consideration it is important only that near two independent $%
\mathbf{K}$ points the dispersion $\xi \equiv |\phi (\mathbf{p})|=\hbar
v_{F}|\mathbf{p}|$, where $v_{F}$ is the Fermi velocity and the the wave
vector $\mathbf{p}$ is measured from the corresponding $\mathbf{K}$ point. In
the result of diagonalization of the operator (\ref{Hamgraph}) one finds that
the presence of the gap $\Delta $ in it breaks the equivalence between $A$ and $B
$ sublattices, and the spectrum of the quasiparticle excitations close to the  $%
\mathbf{K}$ points takes the form $E\left( \mathbf{p}\right) =\pm \sqrt{%
\hbar ^{2}v_{F}^{2}\mathbf{p}^{2}+\Delta ^{2}}-\mu $.

We will account for the
quasiparticle scattering from impurities  in the
framework of the Abrikosov-Gorkov scheme, writing the self-consistent
equation for self-energy in the matrix form
\begin{equation}
\widehat{\Sigma} (\mathbf{p},\varepsilon _{n})=n_{i}\int_{BZ}\frac{d^{2}q}{%
(2\pi )^{2}}\widehat{V}(\mathbf{q})\widehat{G}(\mathbf{p-q},\varepsilon
_{n}) \widehat{V}(\mathbf{q}),  \label{Sigma-eq}
\end{equation}%
with $n_{i}$ as concentration of impurities and $\varepsilon_n = \pi T(2 n+1) $,
the full inverse Green's function (GF)
\begin{equation}
\widehat{G}^{-1}(\mathbf{p},\varepsilon _{n})=\widehat{G}_{0}^{-1}(\mathbf{p}%
,\varepsilon _{n})-\widehat{\Sigma} (\mathbf{p},\varepsilon _{n}),
\label{GF-full}
\end{equation}%
and the free inverse Green's function
\begin{equation}
\widehat{G}_{0}^{-1}(\mathbf{p},\varepsilon _{n})=(i\varepsilon _{n}+\mu )
\widehat{\tau} _{0}-\widehat{\mathcal{H}}(\mathbf{p}).  \label{GF-0}
\end{equation}%
The sublattices $A$ and $B$ are shifted by a distance of the order of
lattice constant $a$, hence their images in inverse space are separated by
momenta of the order of $\hbar /a.$ Hence, for the relatively long-range
potential $\widehat{V}(\mathbf{q})$ in Eq.~(\ref{Sigma-eq}) one can ignore
the quasi-particle scattering between the inequivalent valleys. At the same
time we will assume $\widehat{V}(\mathbf{q})$ as momentum independent for
the intra-valley scattering, i.e.
\begin{equation}
\widehat{V}(\mathbf{q})= \widehat{\tau}_{0}\left\{
\begin{array}{cc}
u(\mathbf{0}), & |q|\lesssim \frac{\max \left\{ |\mu |,\Delta \right\} }{%
\hbar v_{F}} \\
0, & \frac{\max \left\{ |\mu |,\Delta \right\} }{\hbar v_{F}}\ll |q|\lesssim
\frac{\hbar }{a}.%
\end{array}%
\right.  \label{potential}
\end{equation}

The next step in the solution of Eq.~(\ref{Sigma-eq}) is the decomposition of the
self-energy over Pauli matrices $\widehat{\Sigma }(\mathbf{p},\varepsilon
_{n})=\sum_{i=0}^{3}\sigma _{i}(\mathbf{p},\varepsilon _{n})\widehat{\tau }%
_{i}$. One can see that by ignoring the self-consistence procedure in Eq.~(\ref%
{Sigma-eq}) with potential (\ref{potential}) (which does not mix valleys),
one obtains $\widehat{\Sigma }^{\left( 0\right) }(\mathbf{p},\varepsilon
_{n})$ in the diagonal form. It is possible to show that the off-diagonal
components appear in the order $n_{i}^{2}$ only, i.e. $\sigma _{1}^{R}(%
\mathbf{p},\varepsilon _{n})$ and $\sigma _{2}^{R}(\mathbf{p},\varepsilon
_{n})$ terms can be omitted. Finally, the momentum dependence of $\widehat{%
\Sigma }(\mathbf{p},\varepsilon _{n})$ can be also ignored in view of the
constancy of potential (\ref{potential}) in the domain of each valley and
taking into account that the main contribution to $\widehat{\Sigma }$
appears from the relatively large momenta $|\mathbf{q}|\sim \max \left\{
|\mu |,\Delta \right\} /\hbar v_{F}$. This latter fact justifies our use of
the Abrikosov-Gorkov technique for averaging of the quasiparticle scattering
over impurity positions. As a result, after the analytical continuation $%
i\varepsilon _{n}\rightarrow \varepsilon $, the matrix
Eq.~(\ref{Sigma-eq}) reduces to the system of two equations \cite{footnote}
\begin{equation}
\begin{split}
&  \left\{
\begin{array}{c}
\sigma _{0}^{R}(\varepsilon ) \\
\sigma _{3}^{R}(\varepsilon ) \\
\end{array}%
\right\} =\frac{4\hbar }{\pi \tau _{0}|\mu |}\left\{
\begin{array}{c}
\varepsilon +\mu -\sigma _{0}^{R}(\varepsilon ) \\
\Delta +\sigma _{3}^{R}(\varepsilon ) \\
\end{array}%
\right\}  \\
& \times \int_{0}^{W}\frac{\xi d\xi }{\left[ \varepsilon +\mu -\sigma
_{0}^{R}(\varepsilon )\right] ^{2}-\xi ^{2}-[\Delta +\sigma
_{3}^{R}(\varepsilon )]^{2}},
\end{split}
\label{Sigma}
\end{equation}%
where we introduce the \textquotedblleft relaxation time scale\textquotedblright  \cite{Peres2007PRB}
\begin{equation}
\frac{1}{\tau _{0}}= \frac{ n_{i}|u(\mathbf{0})|^{2}|\mu |}{4\hbar ^{3}v_{F}^{2}}.
\label{tau0}
\end{equation}

The real parts $\mbox{Re}\, \sigma _{0}^R$ and $\mbox{Re}\, \sigma _{3}^R$,
which are logarithmically dependent on the high energy cutoff $W,$ can be
included in the renormalized $\mu $ and $\Delta $, respectively. For the
self-energy imaginary parts, which determine the quasiparticle relaxation
rate, one finds
\begin{equation}  \label{sigmafinal}
\begin{split}
& \! \! \! \! \! \! \! \! \!
\left\{ \! \!
\begin{array}{c}
\mbox{Im} \, \sigma _{0}^{R}( \varepsilon) \\
\mbox{Im} \, \sigma _{3}^{R}( \varepsilon ) \\
\end{array}
\! \! \right\} = - \frac{2 \theta \left[ \left( \varepsilon +\mu \right)^{2}-\Delta
^{2}\right] \mathrm{sgn}\!\left( \varepsilon+\mu\right)} {(\tau_0/\hbar)
|\mu | } \left\{ \! \!
\begin{array}{c}
\varepsilon+\mu \\
\Delta \\
\end{array}
\! \! \right\}.
\end{split}%
\end{equation}

One can note that the denominator of the full GF (\ref{GF-full}) can be
approximated as:%
\begin{equation*}
\begin{split}
& \lbrack \varepsilon +\mu -i\mbox{Im}\,\sigma _{0}^{R}(\varepsilon )]^{2}-\xi
^{2}-[\Delta +i\mbox{Im}\,\sigma _{3}^{R}(\varepsilon )]^{2} \\
& \approx [\varepsilon +\mu +i\Gamma (\varepsilon )]^{2}-E^{2}(\mathbf{p}),
\end{split}%
\end{equation*}
so that the full inverse GF can now be written as
$[\widehat{G}^{R}(\mathbf{p},\varepsilon + i 0)]^{-1} \approx [\widehat{G}_{0}^{R}(\mathbf{p},\varepsilon)]^{-1} +
i \widehat{\tau}_{0} \Gamma(\varepsilon)$. Here the energy-dependent
scattering rate $\Gamma (\varepsilon )$, central for our consideration,
explicitly appears:
\begin{equation}  \label{Gamma-full}
\begin{split}
\Gamma (\varepsilon )& =-\mbox{Im} \, \sigma _{0}^{R}(\varepsilon )-\frac{\Delta
}{\varepsilon +\mu } \mbox{Im} \, \sigma _{3}^{R}(\varepsilon) \\
 =&\Gamma _{0}\left[ \frac{|\varepsilon +\mu |} {|\mu |}+\frac{\Delta ^{2}}{%
|\varepsilon +\mu | |\mu|}\right] \theta \left[ \left( \varepsilon +\mu
\right) ^{2}-\Delta ^{2}\right]
\end{split}%
\end{equation}%
with $\Gamma _{0}=2 \hbar /\tau_0$. In the numerical results presented below
we use the value $\Gamma_0 = 20 \, \mbox{K}$ ignoring its dependence on the
carrier concentration.

It follows from Eq.~(\ref{Gamma-full}) that for $(\varepsilon +\mu )^2 <
\Delta^2$ the scattering is absent. Further consideration shows that in
spite of the presence of $\Gamma(\varepsilon)$ in the denominators of Eq.~(%
\ref{A-def-new}), this fact does not result in divergence of the physical
observables. Yet, one should keep in mind that some scattering processes
beyond our model along with the next order corrections to the solution (\ref%
{sigmafinal}) can make the scattering rate finite below the gap edge. For
our numerical work we took this into account by adding a small residual
scattering rate $\gamma_0$ to $\Gamma(\varepsilon)$. In accordance to the
theoretical analysis the final results turn out to be practically
independent of the value $\gamma_0$.

Using the Kubo relations one can derive electric conductivity and
thermoelectric coefficient in the explicit form
\begin{equation}  \label{Kubo}
\left\{
\begin{array}{c}
\sigma \\
\beta \\
\end{array}%
\right\} =\frac{e^{2}}{\hbar }\int_{-\infty }^{\infty }\frac{d\varepsilon
\mathcal{A}(\varepsilon ,\Gamma (\varepsilon ),\Delta )}{2T\cosh ^{2}\frac{%
\varepsilon }{2T}}\left\{
\begin{array}{c}
1 \\
\varepsilon /(eT) \\
\end{array}%
\right\} ,
\end{equation}
where in the presence of $\Delta $ the function $\mathcal{A}$ is given by
\cite{Gorbar2002PRB,Gusynin2005PRB}
\begin{equation}
\begin{split}
\mathcal{A}(\varepsilon ,\Gamma & (\varepsilon ),\Delta )=\frac{1}{2\pi ^{2}}%
\left[ 1+\frac{(\mu +\varepsilon )^{2}-\Delta ^{2}+\Gamma ^{2}(\varepsilon )%
}{2|\mu +\varepsilon |\Gamma (\varepsilon )}\right. \\
& \times \left. \left( \frac{\pi }{2}-\arctan \frac{\Delta ^{2}+\Gamma
^{2}(\varepsilon )-(\mu +\varepsilon )^{2}}{2|\mu +\varepsilon |\Gamma
(\varepsilon )}\right) \right] .
\end{split}
\label{A-def-new}
\end{equation}%
For $\Delta=0$ Eq.~(\ref{A-def-new}) reduces to the commonly used expression
(see, e.g., Refs.~\onlinecite{Lofwander2007PRB,Ugarte2011PRB}). In this case, setting
also $\Gamma(\varepsilon) = \Gamma_0 =\mbox{const}$, one obtains for $|\mu|
\gg T, \Gamma_0$ that $\sigma = e^2 |\mu|/(2 \pi \hbar \Gamma_0)$ and $\beta
= \pi e T \mathrm{sgn} \mu/(6 \hbar \Gamma_0)$, in agreement wit Ref.~\onlinecite%
{Gusynin2005PRB}. Then the value of the thermopower $S = - \beta/\sigma$
turns out to be the same as in the conventional metals, $S =- (\pi^2/3e) T/\mu$,
and coincides with the result obtained directly from the Mott formula (\ref{Mott-formula}).

\begin{figure}[h]
\centering{\includegraphics[width=9.cm]{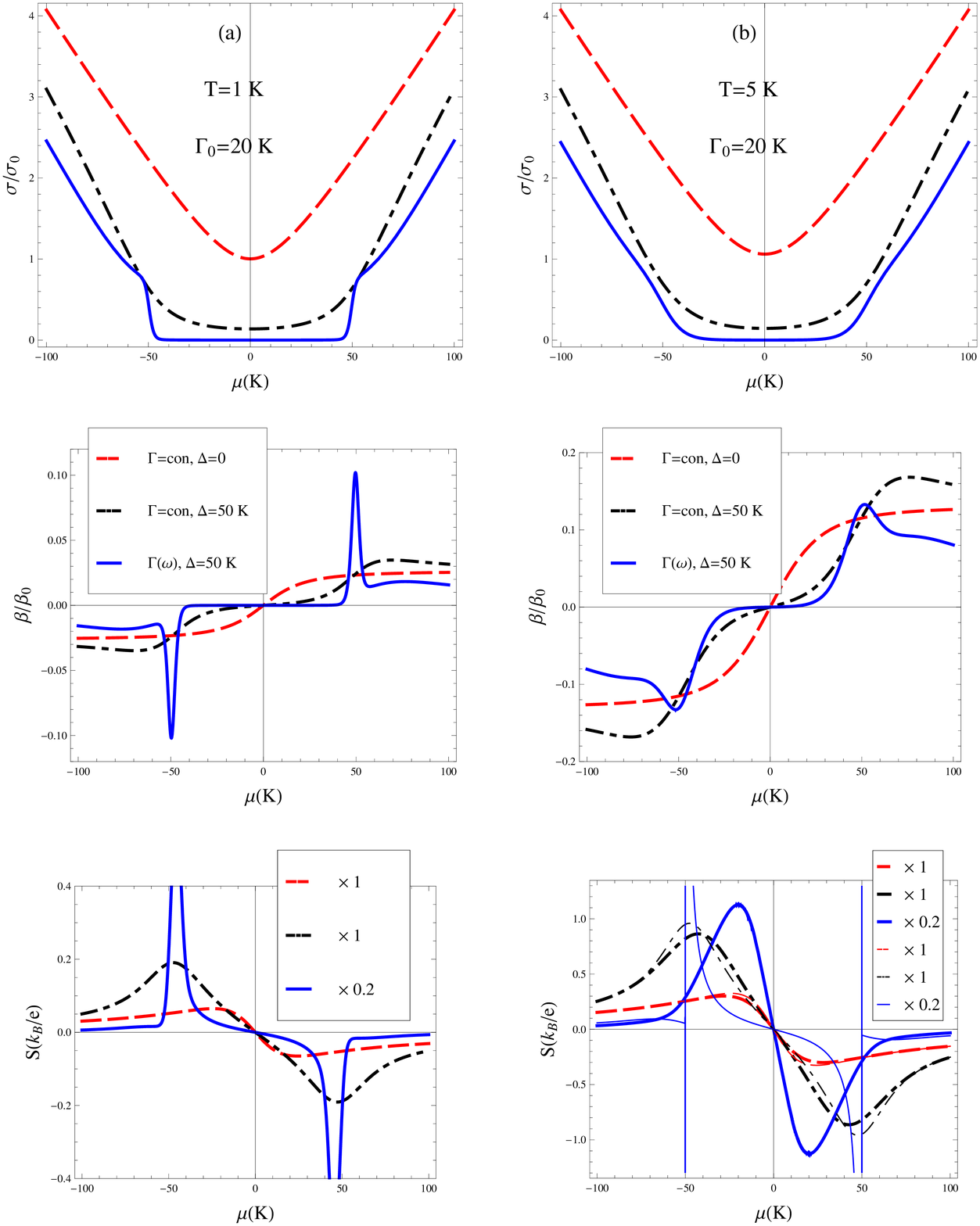}}
\caption{(Color online) Top panel: electrical conductivity $\protect\sigma$
in units of the value $\protect\sigma_0 = 2 e^2/(\protect\pi^2 \hbar)$;
middle panel: thermoelectric coefficient $\protect\beta$ in units of the
value $\protect\beta_0 = k_B e/\hbar$; bottom panel: thermopower $S$ in
units of the value $k_B/e$ as functions of the chemical potential $\protect%
\mu$. Left side: (a) for $T=1 \, \mbox{K}$, and right side: (b) for $T= 5K$.
In all graphs three cases are shown: the dashed (red) curve is for the
energy-independent scattering $\Gamma=\Gamma_0$ and $\Delta = 0$, the
dash-dotted (black) curve is for $\Gamma=\Gamma_0$ and $\Delta = 50 \,
\mbox{K}$, and the solid (blue) curve is for energy-dependent
$\Gamma(\varepsilon)$ and $\Delta = 50 \, \mbox{K}$. The solid curve in the bottom panel
is multiplied by the factor $0.2$. The thin lines in the right bottom panel
are obtained using the Mott formula. }
\label{fig:1}
\end{figure}

The dependences $\sigma(\mu)$, $\beta(\mu)$, and $S(\mu)$ are shown in the
top, middle, and bottom panels of Fig.~\ref{fig:1}, respectively. The left
side [Fig.~\ref{fig:1}~(a)] is for $T=1 \, \mbox{K}$ and the right side
[Fig.~\ref{fig:1}~(b)] is for $T=5 \, \mbox{K}$. The dashed (red) curves in
all panels correspond to the reference case $\Delta=0$, $\Gamma(\varepsilon)
= \Gamma_0$, so that $\sigma(\mu) \varpropto |\mu|$ and $S(\mu) \varpropto
1/\mu$ for large $|\mu|$.
The general expressions (\ref{Kubo}) and (\ref{A-def-new}) also allow one to
reproduce the gapped case with the energy-independent scattering $%
\Gamma(\varepsilon) = \Gamma_0$ \cite{Gorbar2002PRB,Gusynin2005PRB}. The
corresponding reference dependences computed for $\Delta = 50 \, \mbox{K}$
are presented in all panels by the dash-dotted (black) curves.

Our main results obtained with the energy dependent $\Gamma(\varepsilon)$
given by Eq.~(\ref{Gamma-full}) and $\Delta = 50 \, \mbox{K}$ are presented
in all panels by the the solid (blue) curves.

The behavior of the conductivity $\sigma(\mu)$ is shown in the top panel
of Fig.~\ref{fig:1}~(a) and (b). One can see  that it drastically changes due to
account for the energy dependence of $\Gamma(\varepsilon)$.
Namely, a clear kink in the dependence $\sigma(\mu)$ appears
at the gap edge, at $|\mu| = \Delta$, while below it, when $|\mu| < \Delta$,
the value of $\sigma(\mu)$ is strongly suppressed as compared to the case $%
\Gamma(\varepsilon) = \Gamma_0$, $\Delta = \mbox{const}$. This kink is
smeared out with the growth of temperature [Fig.~\ref{fig:1}~(b), $T = 5\, %
\mbox{K}$].

Let us continue to the discussion of the Seebeck signal. In the gapless case
with the constant scattering rate $\Gamma(\varepsilon) = \Gamma_0$ the
signal monotonously changes with $\mu$ passing zero without any visible
anomaly. The gap in the quasiparticle spectrum shows up as
smoothed bumps at $|\mu| = \Delta$, which are rapidly smeared out by
temperature (middle and bottom panels of Fig.~\ref{fig:1}).

The behavior of thermoelectric coefficient $\beta (\mu )$ and thermopower $%
S(\mu )$ in the case under consideration of gapped graphene with 
energy-dependent relaxation time is shown in the middle and bottom panels of
Fig.~\ref{fig:1} by the solid curves. Let us stress that the curves
corresponding to $S(\mu )$ in the bottom panel are multiplied by the factor
0.2 to present them conveniently on the background of the previous cases.
This means that the peaks of the Seebeck signal are at least five times higher
than those ones obtained for $\Gamma (\varepsilon )=\mbox{const}$ case.

A strong enhancement of the Seebeck signal in the case when $%
\Gamma(\varepsilon)$ is energy dependent 
could be foreseen even basing on the Mott formula (\ref{Mott-formula}). 
However, this formula gives
only a hint of the singular behavior of the Seebeck signal and cannot be
used for any quantitative description. Indeed, the thin lines in the bottom
panel of Fig.~\ref{fig:1}~(b) are computed using the zero-temperature
electrical conductivity $\sigma(\mu,T=0) = (2 e^2/\hbar) \mathcal{A}(0
,\Gamma (0 ),\Delta )$ and the Mott formula (\ref{Mott-formula}), while the
thick lines in the middle and bottom panels of Fig.~\ref{fig:1} are plotted
using the Kubo formulas (\ref{Kubo}) both for $\sigma$ and $\beta$. One finds
that for the case of $\Delta =0$ and $\Gamma(\varepsilon) = \mbox{const}$ agreement
between the Kubo and Mott formulas is very good. We checked that for $T =
1\, \mbox{K}$ it becomes perfect, so that the lines for the Mott formula are
not shown in the left bottom panel of Fig.~\ref{fig:1}. The right bottom
panel shows that for the case of finite $\Delta$ and $\Gamma(\varepsilon) = %
\mbox{const}$ one can already see some discrepancies between the Kubo and
Mott formulas, especially near $|\mu| = \Delta$. Finally, the Mott formula
fails completely when the energy dependence of $\Gamma(\varepsilon)$ is
taken into account.

Two more comments on the obtained bump of the Seebeck signal should be made.
First, the shape of the bump depends on the presence of the $\sim \Delta^2$ term in
Eq.~(\ref{Gamma-full}) and thus accounting for the self-energy $%
\sigma_3(\varepsilon)$ is important for the qualitative theory. Second, our
arguments are also directly applicable to gapped bilayer graphene, and
indeed the computations done in Ref.~\onlinecite{Hao2010PRB} confirm this.

The applicability of the model potential (\ref{potential}) to the case under
consideration deserves a more detailed discussion. It is worth stressing
that we use it to solve the equation for self-energy with further fixation
of $\Gamma _{0}=2\hbar \tau _{0}^{-1}\left( |\mu |=\Delta \right) =n_{i}|u(%
\mathbf{0})|^{2}|\Delta |/(2\hbar ^{2}v_{F}^{2})$ [see Eq.~(\ref{tau0})].
This procedure gives a consistent analytical treatment of the problem
close to the gap edge $|\mu -\Delta |\ll \Delta ,$ but it does not allow
one to reproduce correctly the experimentally observed dependences of $\sigma $ and
$S$ on the carrier concentration $n(\varpropto \mu ^{2}\mathrm{sgn}\mu )$
beyond this region. In order to get a better agreement with the experiment
in a wider interval of concentrations one could use the scattering
potential $\widehat{V}(\mathbf{q})$ in the form of a long-range Coulomb one
(see, for instance, Refs.~\onlinecite{Hwang2009PRB,Ugarte2011PRB,Peres2007PRB,Yan2009PRB}).
In such consideration one obtains that at large $n$ the scattering rate $%
\Gamma _{0}\varpropto 1/|\mu |$ (contrary to our reference case, where $%
\Gamma _{0}= \mbox{const}$), which results in the observed linear dependence
$\sigma (n)\varpropto |n|$ (contrary to our $\sigma (n)\varpropto \sqrt{|n|}$).

The specifics of thermopower consists of its sensitivity to the derivative of
the scattering rate. This is why the presence of the step function in  Eq.~(\ref%
{Gamma-full}) produces much stronger effect on the behavior of $S(\mu )$ in
the vicinity of $|\mu |\approx \Delta $ than a relatively slow energy
dependence of $\Gamma_0$ which could appear from the screened
Coulomb potential $\widehat{V}(\mathbf{q})$.
Let us  again call the reader's attention to the evident analogy between the transport
in gapped graphene and that in metal close to the electronic topological transition.
Indeed,  in the vicinity of the critical point $\mu=\mu_{c}$, when the Fermi surface connectivity changes, the  quasiparticle relaxation rate also acquires a contribution
depending on energy in the form of  step function, what generates the well known kinks in conductivity
and peaks in thermopower  \cite{Blanter1994PR}.

One can imagine that when designing future nanoelectronic devices it will be
possible to control their temperature regime using the Peltier cooling effect,
which is also governed by the value of the thermoelectric coefficient.
As was already demonstrated in Ref.~\onlinecite{Wang2011PRL}, thermoelectric power can be tuned by controlling the band gap in  dual-gated bilayer graphene, which looks promising for practical applications.

Turning this around, one can exploit the predicted giant peak of the Seebeck signal as
a signature of the gap opening. Its existence in the quasiparticle spectrum of single layer graphene presents an intersting problem. Although there is not so much evidence \cite%
{Zhou2007NatMat,Li2009PRL} that this gap is present in zero magnetic field,
there is a growing confidence that the $\nu =0$ quantum Hall state in graphene
is gapped (see Ref.~\onlinecite{Zhao2012} and references therein), so that a
generalization of the present work for a finite magnetic may present some
interest.

S.G.Sh. thanks Yu.V. Skrypnyk for useful discussion. This work was supported
by SIMTECH Grant No. 246937 of the European FP7 program. S.G.Sh. was
supported by  SCOPES Grant No. IZ73Z0\verb|_|128026 of the Swiss NSF, by a
grant from the Swedish Institute, and by 
Ukrainian-Russian SFFR-RFBR Grant No. F40.2/108.


\begin{thebibliography}{99}
\bibitem{xx} The subscript $xx$ in the diagonal components $\sigma $ and $S$
is omitted, because in absence of magnetic field the off-diagonal components
are absent.
\end{thebibliography}
\end{document}